# QUANTUM DIMENSIONAL ZEEMAN EFFECT IN THE MAGNETO–OPTICAL ABSORPTION SPECTRUM FOR QUANTUM DOT – IMPURITY CENTER SYSTEMS

**V.D. Krevchik**[1], **A.B. Grunin**[1], **A.K. Aringazin**[2,3], and **M.B. Semenov**[1,3]

[1]Department of Physics, Penza State University,
Penza, 440017, Russia
physics@diamond.stup.ac.ru

[2]Department of Theoretical Physics, Institute for Basic
Research, Eurasian National University, Astana, 473021, Kazakstan

[3]Institute for Basic Research, P.O. Box 1577,
Palm Harbor, FL 34682, USA
ibr@gte.net

## Abstract

Magneto-optical properties of the quantum dot -- impurity center (QD-IC) systems synthesized in a transparent dielectric matrix are considered. For the QD one-electron state description the parabolic model of the confinement potential is used. Within the framework of zero-range potential model and the effective mass approach, the light impurity absorption coefficient for the case of transversal polarization with respect to the applied magnetic field direction, with consideration of the QD size dispersion, has been analytically calculated. It is shown that for the case of transversal polarization the light impurity absorption spectrum is characterized by the quantum dimensional Zeeman effect.

# 1 Introduction

Magneto-optics of the quantum dot - impurity center (QD-IC) system synthesized in a transparent dielectric matrix is of great interest because of the possibility to construct photodetectors with the guided operating frequency and with sensitivity in the light impurity absorption region.

The magneto-optical absorption in the multi-well quantum systems GaAs – $Ga_{0.75}Al_{0.25}As$ with the $D^-$-states participation has been experimentally investigated in [1]. The variational approach to the $D^-$-center electron localized state description is usually used in the experimental data analysis [2]. This approach has some well-known disadvantages, with the most essential among them being selection of the trial wave function. As it was shown earlier [3, 4], the zero-range potential model allows to obtain an analytic solution for the localized carrier wave function and also to analyze the random position effect in the semiconductive quantum well (QW) and quantum dot (QD) with parabolic potential profile without limitation to the number of dimensionally quantized states, which participates in the localized state formation.

The aim of this work is theoretical analysis of the magneto-optical absorption of the QD-IC systems synthesized in a transparent dielectric matrix. We consider the case of light polarization transversal to the direction of magnetic field. The impurity potential is simulated by the zero-range potential characterized by the intensity $\gamma=2\pi/\alpha$ [4]:

$$V_\delta(\vec{r},\vec{R}_a)=\gamma\,\delta(\vec{r}-\vec{R}_a)\left[1+(\vec{r}-\vec{R}_a)\,\vec{\nabla}_{\vec{r}}\right], \tag{1}$$

where $\alpha$ is determined by the electron-binding energy $E_i$ of the localized state on the same impurity center for massive semiconductor; impurity center (IC) is localized at the point $\vec{R}_a = (x_a, y_a, z_a)$.

Such a model, as it is known [5], is appropriate for description of $D^-$-states which correspond to an additional electron connection to the "shallow" donor. As it has been mentioned above, Lippman-Schwinger equation permits an analytic solution for the wave function $\Psi_\lambda\left(\vec{r},\vec{R}_a\right)$ of the electron, which is localized on the QD short-range potential with parabolic profile [4]:

$$\Psi_\lambda(\vec{r},\vec{R}_a) = C\exp\left(-\frac{r^2+R_a^2}{2a^2}\right)\int_0^\infty dt\, e^{-(\varepsilon_a+3/2)t}\left(1-e^{-2t}\right)^{-3/2}\times$$
$$\times\exp\left\{-\frac{e^{-2t}\left(r^2+R_a^2\right)-2e^{-t}(\vec{r},\vec{R}_a)}{a^2\left(1-e^{-2t}\right)}\right\}. \qquad (2)$$

Here, $C=\left[-\frac{\partial}{\partial\varepsilon_a}G(R_a,R_a;\varepsilon_a)\cdot a^3\right]^{1/2}$ is the normalization factor; $\varepsilon_a=|E_\lambda|/(\hbar\omega_0)$; $E_\lambda=-\hbar^2\ddot{e}^2/(2m^*)$ is the impurity center binding energy; $m^*$ is the electron effective mass; $\omega_0$ is the confinement potential characteristic frequency, which is related to QD-radius $R_0$ and QD-potential amplitude $U_0$ due to $2U_0=m^*\omega_0^2R_0^2$.

The equation which determines the impurity center state binding energy $E_\lambda$ dependence on QD-parameters and IC-position $R_a$ is of the following form [4]:

$$\sqrt{\eta^2+\frac{3}{2}\beta^{-1}}=\eta_i-\sqrt{\frac{2}{\beta\pi}}\int_0^\infty dt\exp\left[-\left(\beta\eta^2+3/2\right)t\right]\times$$
$$\times\left[\frac{1}{2t\sqrt{2t}}-\frac{1}{\left(1-\exp(-2t)\right)^{3/2}}\cdot\exp\left\{-\frac{R_a^{*2}\beta^{-1}}{2}\frac{\left[1-\exp(-t)\right]}{\left[1+\exp(-t)\right]}\right\}\right], \qquad (3)$$

where $\eta^2=|E_\lambda|/E_d$ and $\eta_i^2=|E_i|/E_d$, are parameters describing the IC-state binding energy in QD and in massive semiconductor, respectively; $E_d=m^*e^4/(32\pi^2\hbar^2\varepsilon_0^2\varepsilon^2)$is the effective Bohr energy with account of the effective mass m* and dielectric permeability ε; $\beta=R_0^*/\left(4\sqrt{U_0^*}\right)$; $R_0^*=2R_0/a_d$; $U_0^*=U_0/E_d$; and $R_a^*=R_a/a_d$.

For the QD one-electron-state description the confinement potential of the form $V(r)=m^*\omega_0^2r^2/2$ has been used. It should be noted that for the QD one-electron-state theoretical description the "hard sides" model is often used, i.e. the confinement potential is taken as a spherically symmetric potential well with infinite sides. More accurate approach to the holding (or confinement) potential form requires seeking for a self-consistent solution of Poisson and Schrödinger equations. Numerical analysis of the solution for these equations in the quantum well (QW) case shows [6, 12] that the confinement potential is approximately parabolic potential, but with the bottom (or lower potential part) cut-off. Such potential form is similar to a parabolic one, that allows to treat such a potential as quite a realistic variant, under the confinement potential alternative choice. A convenience of the parabolic potential model for theoretical research of optical properties of quasi-0D-structures in magnetic field is motivated by the fact that, as it will be shown below, such a potential allows to obtain analytic expressions for the light impurity absorption coefficients in the longitudinal [13] or transversal polarization case, with an account of the QD size dispersion.

We consider the impurity electron strong localization case, $\lambda a \gg 1$ ($\lambda^2 \equiv 2m^*|E_\lambda|/\hbar^2$, $a = \sqrt{\hbar/(m^*\omega_0)}$). This allows us to consider one-electron states in magnetic field as the states, which are not distorted by the impurity potential. In the nonsymmetrical gauge fixing of the vector-potential, $\vec{A} = [\vec{B}, \vec{r}]/2$, the one-electron states $\Psi_{n1,m,n2}(\rho, \varphi, z)$, which are not affected by the impurities, and the corresponding energies $E_{n1,m,n2}$ can be obtained,

$$\Psi_{n_1,m,n_2}(\rho,\varphi,z) = \frac{1}{a_1^{|m|+1}} \left[ \frac{(n_1 + |m|)!}{2^{n_2+|m|+1} n_1! n_2! \pi \sqrt{\pi} (|m|!)^2 a} \right]^{1/2} \rho^{|m|} \times$$

$$\times\exp\left[-\left(\frac{\rho^2}{4a_1^2}+\frac{z^2}{2a^2}\right)\right]H_{n_2}\left(\frac{z}{a}\right)F\left(-n_1,|m|+1,\frac{\rho^2}{2a_1^2}\right)\exp(im\varphi) \tag{4}$$

$$E_{n_1,m,n_2}=\frac{|e|B\hbar m}{2m^*}+\hbar\omega_0(n_2+1/2)+\hbar\sqrt{\omega_0^2+\frac{e^2B^2}{4m^{*2}}}\left(2n_1+|m|+1\right), \tag{5}$$

where $\rho, \varphi, z$ are cylindrical coordinates; $H_n(x)$ is Hermite polynomials; $F(\alpha, \beta, x)$ is the confluent hypergeometric function [7]; $a_1^2=a^2/(2\sqrt{1+a^4/(4a_B^4)})$; $a_B=\sqrt{\hbar/(m^*\Omega)}$ is the magnetic length; $n_1$, $n_2$ = 0, 1, 2,…are quantum numbers corresponding to Landau levels and to energy levels for spherically-symmetric oscillator potential; $m$ = 0, ±1, ±2,… is magnetic quantum number.

Below we consider the low magnetic field case, at which QD-impurity ground state is not perturbed. This means that $|E_\lambda|+3/2\hbar\omega_0 >> \hbar\Omega$ [1)], where $\Omega = |e|B/m^*$ is cyclotron frequency, $|e|$ is charge of electron, and $B$ is the magnetic induction.

Because of the position disorder effect in quantum dot with impurity centers (IC) [4], the IC-binding energy is a decreasing function of the IC-coordinate. Therefore the above restriction on $\vec{B}$ can be slightly relaxed in the case of impurity situated in the QD center. Then, substituting $\vec{R}_a$=(0,0,0) into (2), and using the integral representation of Whittaker function [7] we obtain [4]

---

[1)] Photo-ionization of deep impurity centers in an external magnetic field for the case of massive semiconductor was investigated theoretically in refs. [8, 9].

$$\psi_{\lambda}(r) = C\left(\frac{r^2}{a^2}\right)^{-3/4} \cdot \Gamma\left(\frac{\varepsilon_a + 3/2}{2}\right) \cdot W_{-\frac{\varepsilon_a}{2},\frac{1}{4}}\left(\frac{r^2}{a^2}\right). \qquad (6)$$

Here Γ(x) is Euler gamma function, $W_{\kappa,\mu}$ (x) is Whittaker function [7],

$$C = \{2\sqrt{\pi}\Gamma(\varepsilon_a/2 + 7/4) \cdot a^3[(\varepsilon_a/2 + 3/4) \times$$

$$\times (\Psi(\varepsilon_a/2 + 7/4) - \Psi(\varepsilon_a/2 + 1/4)) - 1] / [(\varepsilon_a + 3/2)^2 \Gamma(\varepsilon_a/2 + 1/4)]\}^{-1/2}$$

and Ψ(x) is the logarithmic derivative of gamma function [7].

To our best knowledge, no experimental study of the light impurity absorption in QD semiconductive structures in magnetic field has been carried out, but modern delta-doping technology [10] may be capable to provide such a study.

## 2 The absorption coefficient

Let us consider the light absorption by the QD-IC complex in the case when $\vec{B} \perp \vec{e}_{\lambda}$. The effective Hamiltonian for interaction with the light wave field, $\hat{H}_{\text{int}}^{(t)}$, in the case of transversal polarization $\vec{e}_{\lambda t}$ is taken as

$$\hat{H}_{\text{int}}^{(t)} = \lambda_0 \sqrt{\frac{2\pi \hbar^2 \alpha^*}{m^{*2} \omega} I_0}\; e^{i \vec{q} \vec{r}} \left[(\vec{e}_{\lambda t}, \hat{\vec{P}}) - \frac{|e|B}{2}\left[\vec{e}_{\lambda t}, \vec{r}\right]_z\right], \qquad (7)$$

where $\lambda_0$ is the local field factor, $\alpha^*$ is the fine structure constant with account of the dielectric permeability $\varepsilon$, $I_0$ is the light intensity, $\omega$ is light frequency, $q$ is wave vector, and $\hat{\vec{P}}$ is the electron momentum operator.

The matrix element $M_{f\lambda}^{(t)}$ in the dipole approximation for transition can be presented as (see Appendix for details)

$$M_{f\lambda}^{(t)} = \frac{(-1)^n i\pi\lambda_0 \exp(\mp i\vartheta)}{2^{n+1} n! a_1^2} \sqrt{\frac{\alpha^* I_0}{\omega}} \beta_1 a^3 \frac{[(2n)!]^{1/2} (n_1+1)^{1/2} \Gamma\left(\frac{\beta_1}{2}+n\right) \left[\Gamma\left(\frac{\beta_1}{2}-\frac{1}{2}\right)\right]^{1/2}}{\left[\Gamma\left(\frac{\beta_1}{2}+1\right)\right]^{1/2} \left[\frac{\beta_1}{2}\left(\Psi\left(\frac{\beta_1}{2}+1\right)-\Psi\left(\frac{\beta_1}{2}-\frac{1}{2}\right)\right)-1\right]^{1/2}} \times$$

$$\times \left( (E_{n_1,\pm 1,2n} - E_\lambda) \pm \frac{|e|B\hbar}{2m^*} \right) \sum_{k=0}^{n_1} (-1)^k C_{n_1}^k \left(1+\frac{a^4}{4a_B^4}\right)^{k/2} \frac{2^{k+1}\Gamma(k+2)}{\left(1+\sqrt{1+\frac{a^4}{4a_B^4}}\right)^{k+2}} \times$$

$$\times \frac{1}{\Gamma\left(\frac{\beta_1}{2}+n+k+2\right)} F\left( \frac{\beta_1}{2}+n, k+2; \frac{\beta_1}{2}+n+k+2, 1-\frac{2}{1+\sqrt{1+\frac{a^4}{4a_B^4}}} \right), \quad (8)$$

where $\vartheta$ is the polar angle for the transversal polarization vector $\vec{e}_{\lambda t}$ in cylindrical coordinates, $F(\alpha,\beta;\gamma,z)$ is hypergeometric Gauss function [7], $\beta_1 = \varepsilon_a + \frac{3}{2}$, and $C_{n_1}^k$ is binomial coefficient. In this case, selections rules arises due to the following integrals:

$$\int_0^{2\pi}\exp(-im\varphi)\cos(\varphi-\vartheta)d\varphi=\begin{cases}\pi\exp(\mp i\vartheta), & \text{if } \mathrm{m}=\pm 1,\\ 0, & \text{if } \mathrm{m}\neq\pm 1\end{cases},\tag{9}$$

$$\int_0^{2\pi}\exp(-im\varphi)\sin(\varphi-\vartheta)\,d\varphi=\begin{cases}\mp\pi\, i\exp(\mp i\vartheta), & \text{if } \mathrm{m}=\pm 1,\\ 0, & \text{if } \mathrm{m}\neq\pm 1\end{cases},\tag{10}$$

$$\int_{-\infty}^{\infty}\exp\left[-(1+t)\frac{z^2}{a^2}\right]H_{n_2}\left(\frac{z}{a}\right)dz=\begin{cases}0, \text{if } \mathrm{n}_2\neq 2n, \mathrm{n}=0,1,2,...,\\ (-1)^n\sqrt{\pi}a\frac{(2n)!}{n!}t^n(1+t)^{-n-1/2}, \text{if } \mathrm{n}_2=2n.\end{cases}\tag{11}$$

From Eqs. (9)-(11) one can see that optical transitions from the impurity level are possible only to states with the quantum number values $m=\pm 1$ and with even values of quantum number $n_2$. The light impurity absorption coefficient $K^{(t)}(\omega)$ in the case of transversal polarization, with an account of QD sizes dispersion[2)], can be represented as

$$K^{(t)}(\omega)=\frac{2\pi\cdot N_0}{\hbar\cdot I_0}\sum_{n1,n}\sum_{m=-1}^{1}\delta_{|m|,1}\int_0^{3/2}du\,P(u)\left|M_{f\lambda}^{(t)}\right|^2\cdot\frac{1}{\hbar\omega_0\beta^*\left(X-\eta^2\right)}\times$$

---

[2)] It is supposed that the dispersion arises during phase decay processes in resaturated solid solution and has been satisfactorily described by the Lifshits-Slezov formula [11]:

$$P(u=R_0/\overline{R}_0)=\begin{cases}\frac{3^4eu^2\exp[-1/(1-2u/3)]}{2^{5/3}(3+u)^{7/3}(3/2-u)^{11/3}}, & u<3/2\\ 0, & u>3/2\end{cases}$$

where e is the natural logarithm base, $R_0$ and $\overline{R_0}$ are QD-radius and mean value of QD-radius, respectively.

$$\times\delta\left[\frac{(2n+1/2)+\sqrt{1+\beta^{*2}u^2/a^{*4}}\,(2n_1+2)}{\beta^*(X-\eta^2)}+\frac{mu}{a^{*2}(X-\eta^2)}-u\right] \tag{12}$$

where $\delta_{|m|,1}$ is Kronecker symbol,

$$\delta_{|m|,1}=\begin{cases}\mathbf{1}, if\ \ m=\pm\mathbf{1}\\ \mathbf{0}, if\ \ m\neq\pm\mathbf{1},\end{cases}$$

$N_0$ is QD concentration in dielectric matrix; $P(u)$ is Lifshits-Slezov function [11]; $\delta(z)$ is Dirac delta function; $X=\hbar\omega/E_d$ is photon energy in the effective Bohr energy units, $\beta^*=\overline{R_0^*}/(4\sqrt{\overline{U_0^*}})$, $\overline{R_0^*}=2\overline{R_0}/a_d$, and $a^*=a_B/a_d$. Zeros of the argument of Dirac delta function are found from the equation,

$$\frac{(2n+1/2)+(2n_1+2)\sqrt{1+\beta^{*2}u^2/a^{*4}}}{\beta^*(X-\eta^2)}+\frac{mu}{a^{*2}(X-\eta^2)}-u=0 \tag{13}$$

In the case of transversal light polarization, we can find that Eq. (13) has only two roots, $u_{n1,n,\pm1,}$ satisfying the energy conservation law for the optical transition,

$$u_{n1,n,-1}=\frac{\left(2n+\frac{1}{2}\right)\left[1+a^{*2}(X-\eta^2)\right]+(2n_1+2)\sqrt{\left[1+a^{*2}(X-\eta^2)\right]^2+\left(2n+\frac{1}{2}\right)^2-(2n_1+2)^2}}{\frac{\beta^*}{a^{*2}}\left\{\left[1+a^{*2}(X-\eta^2)\right]^2-(2n_1+2)^2\right\}} \tag{14}$$

$$u_{n1,n,+1}=\frac{\left(2n+\frac{1}{2}\right)\left[a^{*2}\left(X-\eta^2\right)-1\right]+\left(2n_1+2\right)\sqrt{\left[a^{*2}\left(X-\eta^2\right)-1\right]^2+\left(2n+\frac{1}{2}\right)^2-\left(2n_1+2\right)^2}}{\frac{\beta^*}{a^{*2}}\left\{\left[a^{*2}\left(X-\eta^2\right)-1\right]^2-\left(2n_1+2\right)^2\right\}}$$

(15)

As the result, we obtain the final expression for light impurity absorption coefficient $K^{(t)}(\omega)$, for the case of transversal polarization,

$$K^{(t)}(\omega)=K_0\beta^*a^{*6}X\left[\left(1-\frac{1}{a^{*2}X}\right)^2\sum_{n_1,n}\frac{(2n)!(n_1+1)(2n_1+2)^2}{2^{2n}(n!)^2}\frac{\left(2c_{n_1,n,-1}+3/2\right)^2}{\Gamma\left(c_{n_1,n,-1}+7/4\right)}\times\right.$$

$$\times\frac{\Gamma\left(c_{n_1,n,-1}+1/4\right)\Gamma^2\left(c_{n_1,n,-1}+3/4+n\right)}{\left[\left(c_{n_1,n,-1}+3/4\right)\left(\Psi\left(c_{n_1,n,-1}+7/4\right)-\Psi\left(c_{n_1,n,-1}+1/4\right)\right)-1\right]}\times$$

$$\times\frac{u^4_{n_1,n,-1}\exp[-1/\left(1-2u_{n_1,n,-1}/3\right)]}{\left(u_{n_1,n,-1}+3\right)^{7/3}\left(3/2-u_{n_1,n,-1}\right)^{11/3}}\times$$

$$\times\left|\beta^*\left[\left(2n_1+2\right)^2-\left(1+a^{*2}\left(X-\eta^2\right)\right)^2\right]u_{n_1,n,-1}+\left(2n+1/2\right)a^{*2}\left(1+a^{*2}\left(X-\eta^2\right)\right)\right|^{-1}\times$$

$$\times\left[\sum_{k=0}^{n_1}(-1)^k C_{n_1}^k \frac{2^{k+1}\Gamma(k+2)}{\Gamma(c_{n_1,n,-1}+11/4+n+k)}\frac{\left[\beta^*\left(1+a^{*2}(X-\eta^2)\right)u_{n_1,n,-1}-(2n+1/2)a^{*2}\right]^{k+\frac{3}{2}}}{\left[\beta^*\left(1+a^{*2}(X-\eta^2)\right)u_{n_1,n,-1}+a^{*2}(2(n_1-n)+3/2)\right]^{k+2}}\times\right.$$

$$\left.\times F\left(c_{n_1,n,-1}+\frac{3}{4}+n,\ k+2;\ c_{n_1,n,-1}+\frac{11}{4}+n+k, 1-\frac{2(2n_1+2)a^{*2}}{\beta^*\left(1+a^{*2}(X-\eta^2)\right)u_{n_1,n,-1}+a^{*2}(2(n_1-n)+3/2)}\right)\right]^2+$$

$$+\left(1+\frac{1}{a^{*2}X}\right)^2\sum_{n_1,n}\frac{(2n)!(n_1+1)(2n_1+2)^2}{2^{2n}(n!)^2}\frac{(2c_{n_1,n,+1}+3/2)^2}{\Gamma(c_{n_1,n,+1}+7/4)}\times$$

$$\times\frac{\Gamma(c_{n_1,n,+1}+1/4)\Gamma^2(c_{n_1,n,+1}+3/4+n)}{\left[(c_{n_1,n,+1}+3/4)\left(\Psi(c_{n_1,n,+1}+7/4)-\Psi(c_{n_1,n,+1}+1/4)\right)-1\right]}\times$$

$$\times\frac{u_{n_1,n,+1}^4\exp[-1/(1-2u_{n_1,n,+1}/3)]}{(u_{n_1,n,+1}+3)^{7/3}(3/2-u_{n_1,n,+1})^{11/3}}\times$$

$$\times\left|\beta^*\left[(2n_1+2)^2-\left(a^{*2}(X-\eta^2)-1\right)^2\right]u_{n_1,n,+1}+(2n+1/2)a^{*2}\left(a^{*2}(X-\eta^2)-1\right)\right|^{-1}\times$$

$$\times\left[\sum_{k=0}^{n_1}(-1)^k C_{n_1}^k \frac{2^{k+1}\Gamma(k+2)}{\Gamma(c_{n_1,n,+1}+11/4+n+k)}\frac{\left[\beta^*\left(a^{*2}(X-\eta^2)-1\right)u_{n_1,n,+1}-(2n+1/2)a^{*2}\right]^{k+\frac{3}{2}}}{\left[\beta^*\left(a^{*2}(X-\eta^2)-1\right)u_{n_1,n,+1}+a^{*2}(2(n_1-n)+3/2)\right]^{k+2}}\times\right.$$

$$\left.\times F\left(c_{n_1,n,+1}+\frac{3}{4}+n,\ k+2;c_{n_1,n,+1}+\frac{11}{4}+n+k, 1-\frac{2(2n_1+2)a^{*2}}{\beta^*\left(a^{*2}(X-\eta^2)-1\right)u_{n_1,n,+1}+a^{*2}(2(n_1-n)+3/2)}\right)\right]^2\Bigg],$$

(16)

where $K_0 = 3^4 2^{1/3} \pi^3 \alpha^* \lambda_0{}^2 e a_d{}^2 N_0$; $c_{n_1,n,\pm 1} = \beta^* \eta^2 u_{n_1,n,\pm 1}/2$.

Fig. 1 shows the light impurity absorption coefficient $K^{(t)}(\omega)$ spectral dependence in the case of the transversal polarization. This dependence has been calculated from Eq. (16) for the optical transition with maximal oscillator force ($m = \pm 1$, $n_1 = n_2 = 0$) in the case of borosilicate glass pigmented by InSb crystallites.

One can see that the impurity absorption band in magnetic field (curve 1) is splitted into Zeeman doublet (curve 2). Also, we observe that the height of absorption peak related to the electron optical transition to the state $m = -1$ is several times smaller then the peak related to the electron optical transition to state with $m = +1$. Such an asymmetry can be understood in view of the process of displacement from the spherically symmetrical potential well for the electron wave function, which corresponds to state with energy $E_{0,-1,0}$. Indeed, since $E_{0,-1,0} < E_0$ ($E_0$ - the QD ground state energy), and due to uncertainty principle, the electron localization radius in this case has to exceed the oscillator characteristic length $a = \sqrt{\hbar/(m^* \omega_0)}$.

## 3 Conclusions

In this paper we have studied magneto-optical absorption by the "quantum dot – impurity center" complexes in a transparent dielectric matrix. For the impurity potential the zero-range potential model has been used, and QD has been described in terms of the parabolic confinement potential model. The QD-potential amplitude $U_0$ is taken as an empirical parameter. We show that for the case of transversal polarization the light impurity absorption spectrum is characterized by the quantum dimensional Zeeman effect.

As in the case of quasi-0D-nanostructures, an external magnetic field can lead to an appreciable lateral geometric confinement that is important in view of the possible design and construction of photodetectors based on two-phase systems with the guided light impurity absorption band.

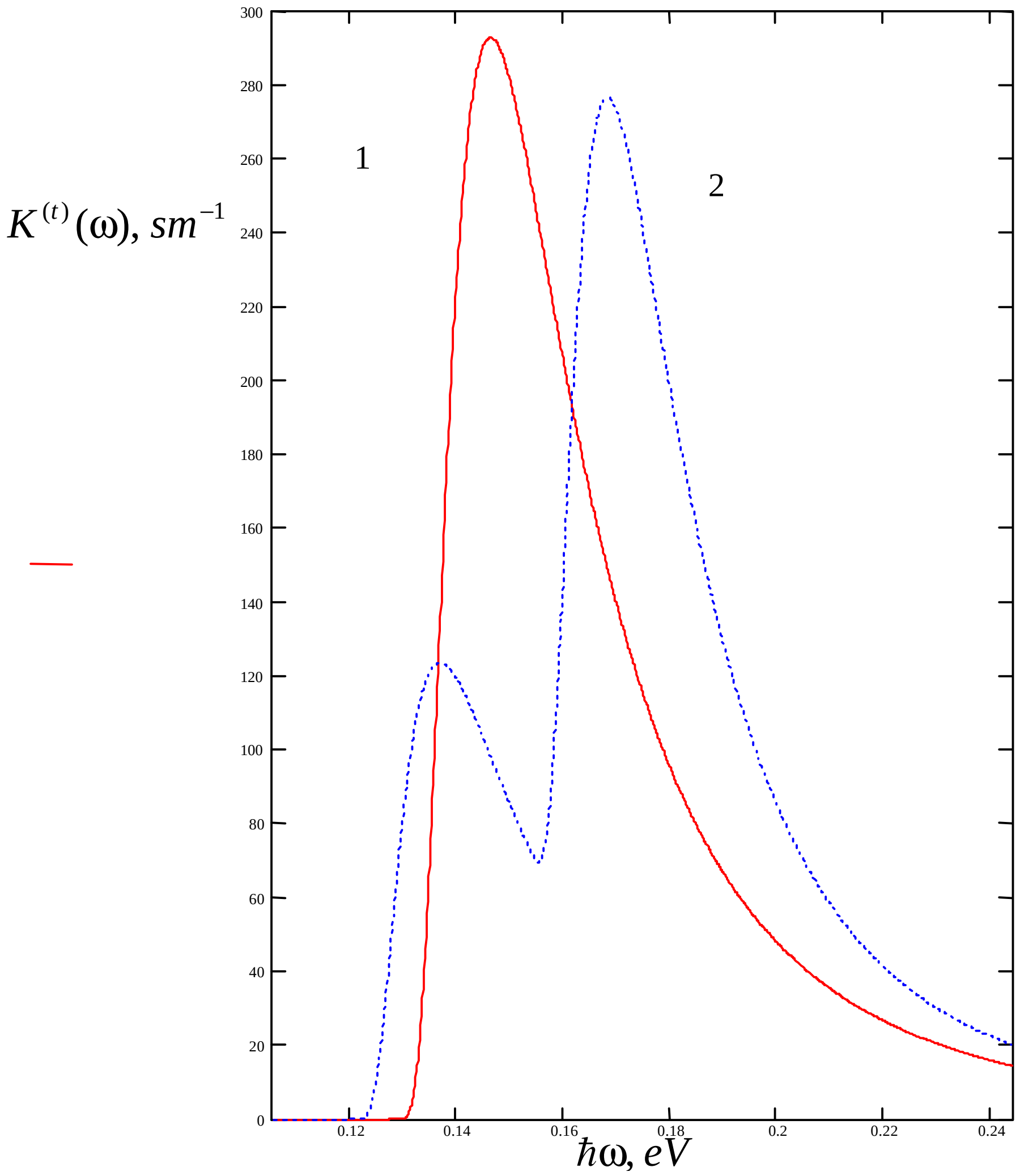

Fig. 1. The light impurity absorption coefficient $K^{(t)}(\omega)$ spectral dependence in the case of the transversal polarization in relation to the magnetic field direction, for the optical transition with maximal oscillator force ($n_1$=0, $n_2$=0, m=±1) in the case of borosilicate glass, which is pigmented by the InSb crystallites ($\overline{R}_0$ =35.9 nm, $U_0$ = 0.2 eV): 1 – $|E_\lambda|$=5.6×10$^{-2}$ eV, B = 0 T; 2 – $|E_\lambda|$=5.6×10$^{-2}$ eV, B = 3.7 T.

# Appendix

The matrix element $M_{f\lambda}^{(t)}$ determines the oscillator force value for the dipole optical transition from the IC (impurity center)-ground state $\Psi_\lambda(r)$ to QD (quantum dot)-discrete spectrum state $\Psi_{n1,m,n2}(\rho, \varphi, z)$ and can be represented as

$$M_{f\lambda}^{(t)} = \lambda_0 \sqrt{\frac{2\pi \hbar^2 \alpha^*}{m^{*2}\omega} I_0} \left\langle \Psi^*_{n1,m,n2} \right| (\vec{e}_{\lambda t}, \hat{\vec{P}}) - \frac{|e|B}{2} [\vec{e}_{\lambda t}, \vec{r}]_z \left| \Psi_\lambda \right\rangle =$$

$$= \lambda_0 \sqrt{\frac{2\pi \hbar^2 \alpha^*}{m^{*2}\omega} I_0}$$

$$\left( \frac{im^*}{\hbar} \left(E_{n1,m,n2} - E_\lambda\right) \left\langle \Psi^*_{n1,m,n2} \right| \left(\vec{e}_{\lambda t}, \vec{r}\right) \left| \Psi_\lambda \right\rangle - \frac{|e|B}{2} \left\langle \Psi^*_{n1,m,n2} \right| \left[\vec{e}_{\lambda t}, \vec{r}\right]_z \left| \Psi_\lambda \right\rangle \right). \quad \text{(A1)}$$

The matrix element $M_{f\lambda}^{(t)}$ in cylindrical coordinates has the following form:

$$M_{f\lambda}^{(t)} = \lambda_0 \sqrt{\frac{2\pi \hbar^2 \alpha^*}{m^{*2}\omega} I_0} \times$$

$$\left( \frac{im^*}{\hbar} \left(E_{n1,m,n2} - E_\lambda\right) \left\langle \Psi^*_{n1,m,n2} \right| \rho \cos(\varphi - \vartheta) \left| \Psi_\lambda \right\rangle - \frac{|e|B}{2} \left\langle \Psi^*_{n1,m,n2} \right| \rho \sin(\varphi - \vartheta) \left| \Psi_\lambda \right\rangle \right)$$

$$= \lambda_0 \sqrt{\frac{2\pi \hbar^2 \alpha^*}{m^{*2}\omega} I_0} \times$$

$$C_{n1,m,n2} C \Gamma\left( \frac{\varepsilon_a + 3/2}{2} \right) \times \int_0^{2\pi} \int_{-\infty}^{\infty} \int_0^{\infty} \rho^{|m|+2} \left( \frac{\rho^2 + z^2}{a^2} \right)^{-3/4} exp\left( -\frac{\rho^2}{4a_1^2} \right) \times$$

$$\times F\left(-n_1,|m|+1,\frac{\rho^2}{2a_1^2}\right)W_{-\frac{\varepsilon_a}{2},\frac{1}{4}}\left(\frac{\rho^2+z^2}{a^2}\right)exp\left(-\frac{z^2}{2a^2}\right)H_{n_2}\left(\frac{z}{a}\right)\times$$

$$\times\left[\frac{im^*}{\hbar}\left(E_{n1,m,n2}-E_\lambda\right)\cos(\varphi-\vartheta)-\frac{|e|B}{2}\sin(\varphi-\vartheta)\right]\exp(-im\varphi)\,d\rho\,d\varphi\,dz, \quad \text{(A2)}$$

where $C_{n1,m,n2}$ and $C$ are the normalization factors for the wave functions $\Psi_{n1,m,n2}(\rho,\varphi,z)$ and $\Psi_\lambda(\rho,z)$, respectively, $\vartheta$ is the polar angle for transversal polarization vector $\vec{e}_{\lambda t}$ in cylindrical coordinates. Integrating over $\varphi$ we obtain

$$\int_0^{2\pi}\left[\frac{im^*}{\hbar}\left(E_{n1,m,n2}-E_\lambda\right)\cos(\varphi-\vartheta)-\frac{|e|B}{2}\sin(\varphi-\vartheta)\right]\exp(-im\varphi)d\varphi=$$

$$=\begin{cases}\frac{im^*}{\hbar}\pi\exp(\mp i\vartheta)\left[E_{n1,\pm1,n2}-E_\lambda\pm\frac{|e|B\hbar}{2m^*}\right], if\ m=\pm1\\ 0, if\ m\neq\pm1\end{cases}$$

(A3)

Using definition of Laguerre polynomial [14, 15] and Whittaker function $W_{-\frac{\varepsilon_a}{2},\frac{1}{4}}\left(\frac{\rho^2+z^2}{a^2}\right)$ integral representation

$$W_{-\frac{\varepsilon_a}{2},\frac{1}{4}}\left(\frac{\rho^2+z^2}{a^2}\right)=\frac{\left(\frac{\rho^2+z^2}{a^2}\right)^{3/4}exp\left(-\frac{\rho^2+z^2}{2a^2}\right)}{\Gamma\left(\frac{\varepsilon_a}{2}+\frac{3}{4}\right)}\int_0^\infty exp\left[-\frac{\left(\rho^2+z^2\right)}{a^2}t\right]t^{\frac{\varepsilon_a}{2}-\frac{1}{4}}(1+t)^{-\frac{\varepsilon_a}{2}-\frac{1}{4}}dt,$$

(A4)

integrating over $\rho$ can be made. Using Eq. (A4) we obtain

$$\int_0^\infty \rho^3\, exp\left[-\frac{\rho^2}{4a_1{}^2}\right] F\left(-n_1, 2, \frac{\rho^2}{2a_1{}^2}\right)\left(\frac{\rho^2+z^2}{a^2}\right)^{-3/4} W_{-\frac{\varepsilon_a}{2},\frac{1}{4}}\left(\frac{\rho^2+z^2}{a^2}\right) d\rho \;=$$

$$= \frac{a^4}{(n_1+1)\,\Gamma\left(\frac{\varepsilon_a}{2}+\frac{3}{4}\right)} \int_0^\infty dt\, t^{\frac{\varepsilon_a}{2}-\frac{1}{4}} (1+t)^{-\frac{\varepsilon_a}{2}-\frac{1}{4}}\, exp\left[-\frac{z^2}{a^2}\left(t+\frac{1}{2}\right)\right] \times$$

$$\times \sum_{k=0}^{n_1} \frac{(-1)^k}{k!}\, \frac{2^{k+1}\,\Gamma(n_1+2)}{\Gamma(n_1-k+1)}\, \frac{\left(1+\frac{a^4}{4a_B{}^4}\right)^{k/2}}{\left(1+\sqrt{1+\frac{a^4}{4a_B{}^4}}+2t\right)^{k+2}} . \qquad \text{(A5)}$$

After integrating over $z$-coordinate, with an account of Eqs. (A5) and (A6) [14, 15],

$$\int_{-\infty}^{\infty} e^{-x^2}\; H_{2n}(xy)\, dx = \sqrt{\pi}\, \frac{(2n)!}{n!}\, \left(y^2-1\right)^n , \qquad \text{(A6)}$$

only $t$-variable integrating remains in the matrix element $\mathrm{M}_{f\,\lambda}^{(t)}$ expression. This integrating can be performed by using the hypergeometric Gauss function integral representation [15],

$$F(a,b;\, a+b+1-c, 1-z) = \frac{\Gamma(a+b+1-c)}{\Gamma(a)\,\Gamma(b+1-c)} \int_0^\infty t^{a-1}\, (1+t)^{c-a-1} (1+zt)^{-b}\; dt\, .$$

(A7)